\documentclass[%
 reprint,
 superscriptaddress,
 amsmath,amssymb,
 aps,
]{revtex4-2}

\usepackage{graphicx}
\usepackage{dcolumn}
\usepackage{bm}
\usepackage{hyperref}
\usepackage{physics}
\usepackage{xcolor}
\usepackage{mathtools}
\usepackage{amssymb}
\usepackage{color,soul}
\usepackage{float}
\usepackage{siunitx}
\usepackage{bbm}
\usepackage{mathtools}

\usepackage{accents}

\newcommand\varpm{\mathbin{\vcenter{\hbox{%
				\oalign{\hfil$\scriptstyle+$\hfil\cr
					\noalign{\kern-.3ex}
					$\scriptscriptstyle({-})$\cr}%
}}}}
\newcommand\varmp{\mathbin{\vcenter{\hbox{%
				\oalign{$\scriptstyle({+})$\cr
					\noalign{\kern-.3ex}
					\hfil$\scriptscriptstyle-$\hfil\cr}%
}}}}

\begin{document}


\title{Geometrical control of topology with orbital angular momentum modes}

\author{Yunjia Zhai}
\email{yunjia.zhai@autonoma.cat}
\affiliation{Departament de F\'{i}sica, Universitat Aut\`{o}noma de Barcelona, 08193 Bellaterra, Spain}

\author{Anselmo M. Marques}
\affiliation{Department of Physics and i3N, University of Aveiro, 3810-193 Aveiro, Portugal}

\author{Ricardo G. Dias}
\affiliation{Department of Physics and i3N, University of Aveiro, 3810-193 Aveiro, Portugal}

\author{Ver\`{o}nica Ahufinger}
\affiliation{Departament de F\'{i}sica, Universitat Aut\`{o}noma de Barcelona, 08193 Bellaterra, Spain}
\affiliation{ICFO -- Institut de Ciencies Fotoniques, The Barcelona Institute of Science and Technology, 08860 Castelldefels, Spain}

\author{David Viedma}
\affiliation{Departament de F\'{i}sica, Universitat Aut\`{o}noma de Barcelona, 08193 Bellaterra, Spain}







\begin{abstract}
We study how the topological properties of a one-dimensional staggered lattice, loaded into states with orbital angular momentum $l=1$, can be controlled simply by tuning the relative angle between sites. The original system under consideration can be depicted as a Creutz ladder model when unwrapping the different state circulations in a synthetic dimension. Depending on the hopping strengths of the chain, different topological regimes may be accessed by changing the ladder angle, as determined by the value of the winding number of the chain. We analytically and numerically explore the different available regimes, and determine the number of topologically protected edge states that exist in each case. We also study the emergence of band inversion across topological transitions and show that it agrees with the winding number calculations, thus serving as an additional topological marker. Then, we propose a realistic experimental implementation in a photonic waveguide system, where the topological transition manifests as a sudden change of the behavior of the propagation of light in the system.
\end{abstract}

\maketitle


\section{Introduction}


Topological insulators (TIs) have become common subjects of study in recent years, and have been implemented in a wide array of physical platforms, including ultracold atoms \cite{Goldman2016,Cooper2019}, photonic \cite{Lu2014,Ozawa2019,Leykam2020} and acoustic systems \cite{Yang2015,
Xue2022}, electrical circuits \cite{Imhof2018,Helbig2020}, and many more. Systems with nontrivial topology display edge or surface states that live in energy gaps, and which are endowed with protection by a symmetry against certain types of disorder and fabrication imperfections, making them of high interest for many applications \cite{Hasan2010,Ozawa2019,Qi2011,deLeseleuc2019,St-Jean2017,Barik2018}.
This has therefore prompted efforts to control the topological phase of any given system, which can be achieved, for instance, by applying external fields \cite{Klitzing1980,Laughlin1981,Thouless1982,Tsui1982,Tome2021,Bhandari2025,Xie2025}, mechanical strain \cite{Allein2023,Inbar2023,Sanchez2025}, Supersymmetry (SUSY) transformations \cite{Miri2013A,Queralto2020,Viedma2022} or by using nonlinearities \cite{Maczewsky2020,Xia2021,Dai2024}, among others. For neutral particles, the engineering of artificial gauge fields (AGFs) \cite{Jaksch2003,Lin2009,Dalibard2011,Struck2012,Fang2012,Aidelsburger2013,Hafezi2013} has become a primary way to access these different topological phases. In lattice systems, AGFs generate a nonzero phase factor in the coupling terms that can simulate a finite flux through closed regions of the lattice, either directly or by employing Floquet driving \cite{Goldman2014b}.

Among the several methods to generate AGFs in lattices, we focus on the usage of orbital angular momentum (OAM) states. The couplings in such a scenario naturally become complex due to the unequal phase windings of states with different OAM charge. Crucially, the generated synthetic flux in the system depends solely on the geometry of the lattice \cite{Polo2016}, and thus can be tuned with precision.
OAM states have been used both in ultracold atoms \cite{Nicolau2024,Polo2016,Pelegri2019,Pelegri2019b,Nicolau2023,Corman2014,Wright2013,Andersen2006} and in photonic systems \cite{Jorg2020,Jiang2023,Wang2024,Viedma2024b}.
In this work, we consider a zig-zag chain structure, with each site hosting OAM states of charge $l=1$.
By splitting these into a synthetic dimension \cite{Celi2014,Ozawa2016,Yuan2018,yu2025}, the chain is formally equivalent to the Creutz ladder model \cite{Creutz2001,Kuno2020,Zurita2020,Nicolau2023,zhai2025} with complex diagonal couplings, where the phase depends on the relative angle of the chain. This may be tuned in conjunction with the distance between sites to explore different phases of the model. We show how, depending on the coupling values, a critical angle separates distinct topological phases. Hence, a topological phase transition that changes the number of protected edge states in the system may be induced simply by sweeping the angle across its critical value. To prove the existence of this transition, we calculate the winding number of the bulk Hamiltonian, and find that it changes as the angle is swept. We also obtain the energy spectrum by exact diagonalization of a finite chain, finding that the number of topological edge states is fully consistent with the bulk–edge correspondence. Furthermore, we analyze the band inversion across the gap-closing point, which reflects the exchange of band parity for the expectation value of the sublattice-exchange operator, and whose appearance is connected with a change in the topological invariant~\cite{Hasan2010,Liu2020b,Widmann2025}.

We then propose an implementation of the system using coupled cylindrical optical waveguides, for which OAM modes form a natural basis. We first simulate the spectrum of modes in different topological phases, and then study the propagation of light in a device where the angle is slowly modified along the propagation direction and the system transitions from a nontrivial to a trivial topological phase. In this scenario, light in the edge state, initially localized, is dispersed into the bulk as the angle crosses its critical value because of the disappearance of the edge states in the spectrum.

This work is organized as follows:
In Section~\ref{sec:physicalsystem}, we first introduce the staggered lattice structure and theoretically explore the properties of the Creutz ladder, which emerges when OAM circulations are considered as a synthetic dimension.
Then, in Section~\ref{sec:TI} we investigate the winding number numerically and propose an analytical method to predict whether the critical angle associated to the topological phase transition exists. We calculate its value for different cases and determine the topological phases that it joins depending on the coupling strengths. We also introduce the framework to analyze the band inversion across topological transitions.
After that, we present some cases of interest in Section~\ref{sec:boson}, analyzing first the scenarios for which the topology is sensitive to the angle of the chain. Afterwards, we show that the topological invariant does not depend on $\phi$ in the limit of long distances between sites. 
This is then demonstrated by finite-element simulations in the proposed waveguide setup in Section~\ref{sec:waveguides}. 
We finish the manuscript by presenting our conclusions in Section~\ref{sec:conclusions}.

\section{Physical system} \label{sec:physicalsystem}
\begin{figure}[h]
    \includegraphics[width=1.0\linewidth]{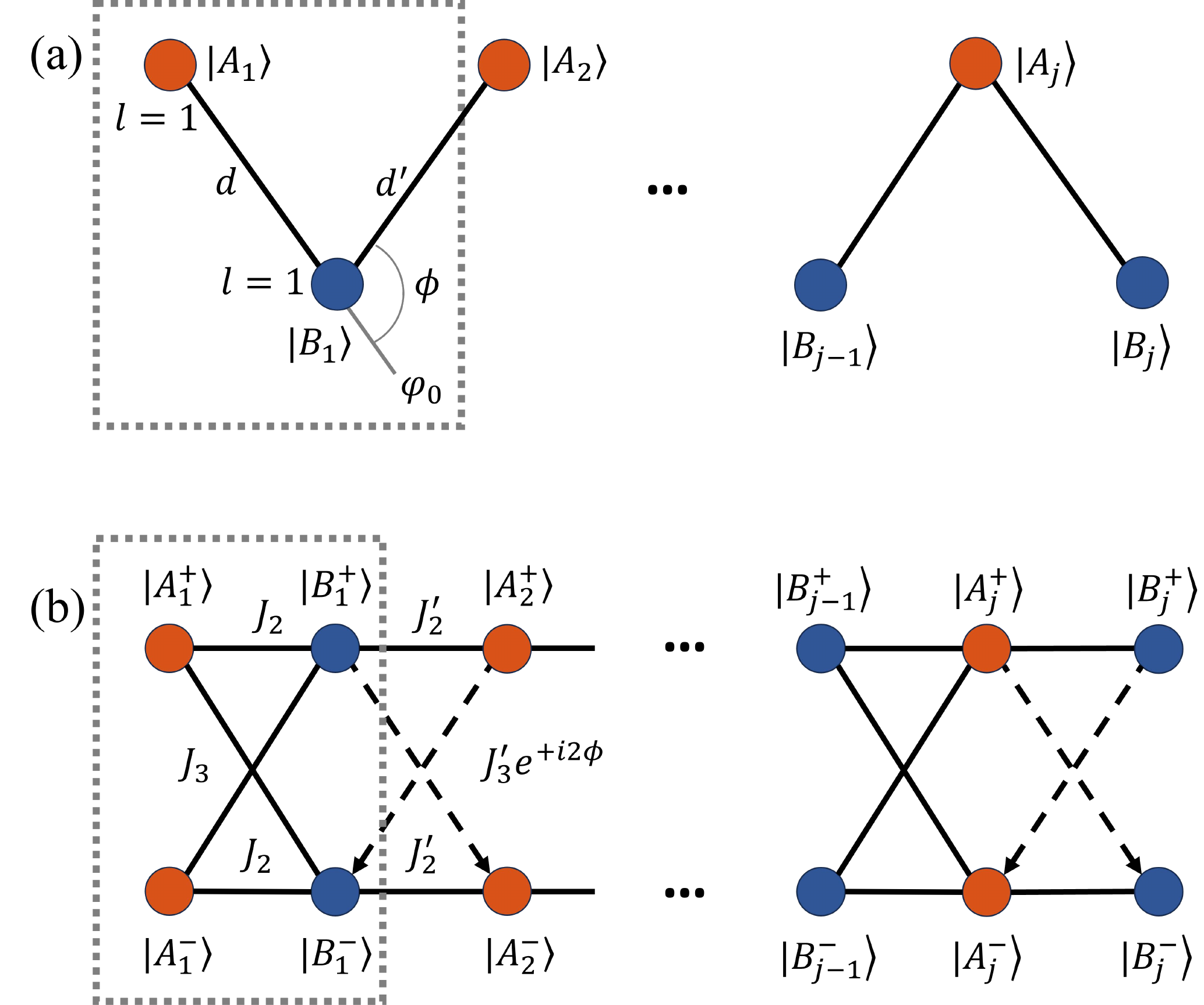}
    \caption{(a) Sketch of the 1D staggered chain of sites loaded with $l=1$ OAM states, with the unit cell marked with the dashed rectangle displaying two sites $A$ and $B$, and with intracell and intercell distances $d$ and $d'$, respectively. The gray line marks the origin of the phase $\varphi_0$. (b) Sketch of the Creutz ladder formed by unwrapping the synthetic dimension spanned by the two circulations of the $l=1$ states in each site $A_j$ and $B_j$, denoted $A_j^{\pm}$ and $B_j^{\pm}$. In this effective model, the unit cell has four sites. We indicate the real tunneling amplitudes with solid lines, and the complex tunneling amplitudes between states with different circulations with dashed lines.}
    \label{fig:l1model}
\end{figure}

We consider a one-dimensional (1D) staggered lattice of sites loaded with OAM states of charge $l=1$, as shown in Fig.~\ref{fig:l1model}(a). Each unit cell $j$ consists of two sites, $A_j$ and $B_j$. The number of unit cells is $N_c$, with a total number of sites of $M=2N_c$. The intracell and intercell separations are given by $d$ and $d'$, and the line that joins sites $B_j$ and $A_{j+1}$ makes an angle $\phi$ relative to the intracell link between $A_j$ and $B_j$.
In this system, the two OAM circulations have the same energy and can be split into different sites in a synthetic dimension \cite{Celi2014}, yielding the unwrapped coupling structure shown in Fig.~\ref{fig:l1model}(b) that can be identified as a Creutz ladder. 
The single-particle Hamiltonian of such a model reads
\begin{align}
    \hat{\mathcal{H}}_{l=1}^0
    &=
    \sum_{\alpha=\pm}
    \bigg(
    J_2 
    \sum_{j=1}^{N_c}
    \hat{a}_j^{\alpha\dagger} 
    \hat{b}_j^{\alpha}
    +
    J'_2 
    \sum_{j=1}^{N_c-1}
    \hat{b}_j^{\alpha\dagger} 
    \hat{a}_{j+1}^{\alpha}
    \nonumber\\
    &\quad
    +
    J_3
    \sum_{j=1}^{N_c}
    \hat{a}_j^{\alpha\dagger} 
    \hat{b}_j^{-\alpha}
    +
    J'_3 e^{-i2\alpha\phi}
    \sum_{j=1}^{N_c-1}
    \hat{b}_j^{\alpha\dagger} 
    \hat{a}_{j+1}^{-\alpha}
    \bigg)
    \nonumber\\
    &\quad
    +\text{H.c.} \label{eq:spHamiltonian}
    ,
\end{align}
where $\hat{a}_j^{\alpha }$ $(\hat{a}_j^{\alpha \dagger})$ and $\hat{b}_j^{\alpha }$ $(\hat{b}_j^{\alpha \dagger})$ are the annihilation (creation) operators of state $\ket{m_j^{\alpha}}$. Here, $j$ is the unit cell index, $m=A,B$ is the site, and $\alpha=\pm$ is the OAM circulation. The coefficient $J_2$~$(J_2^\prime)$ denotes the intracell (intercell) hopping strength between states with the same circulation. In contrast, the coefficient $J_3~(J_3^\prime)$ represents the intracell (intercell) hopping strength between states with opposite circulations. All hopping coefficients in the system are considered to be real.
As shown in Refs.~\cite{Pelegri2019,Viedma2024b,Marques2024}, the amplitudes of all hopping terms $J_2$~$(J'_2)$ and $J_3$~$(J'_3)$ decrease exponentially with distance. Therefore, for $\phi\in(0,\pi/2)$, the next-nearest neighbor hopping between $A_j$~$(B_j)$ and $A_{j+1}$~$(B_{j+1})$ can be safely neglected. The origin of the complex tunneling phases is the azimuthal phase of the OAM states on each site, which appears in the calculation of the overlap integral between wavefunctions corresponding to local eigenstates of neighbouring rings~\cite{Pelegri2019}.
In the considered structure, we fix the phase origin along the line connecting $A_j$ and $B_j$, so that the intracell couplings between the sites are real. The intercell hopping $J'_3$ between $B_j$ and $A_{j+1}$ is offset by an angle $\phi$, which causes the associated phase factor $e^{\pm i2\phi}$.
The topological properties in the limit cases when $\phi=0,~\pi/2$ have been investigated previously \cite{Nicolau2023,Eloi2023}. In those limits, the model can be decomposed into two decoupled Su–Schrieffer–Heeger chains \cite{Su1979}. In contrast, the intermediate region of fluxes displays a richer topological landscape where transitions between different topological phases can occur simply by tuning the angle in Fig.~\ref{fig:l1model}(a). Thus, we consider the range of $\phi\in(0,\pi/2)$ in the following discussion.

\par 

For a chain with $N_c$ unit cells, the momentum-space operators are
\begin{equation}
    \hat a_{k}^{\alpha}=\frac{1}{\sqrt{N_c}}\sum_{j}e^{-ik r_j}\hat a_{j}^{\alpha}, \quad
    \hat b_{k}^{\alpha} = \frac{1}{\sqrt{N_c}}\sum_{j}e^{-ik r_j}\,\hat b_{j}^{\alpha},
\end{equation}
where $r_j$ is the position of the unit cell $j$ and the lattice constant is set to $a_{\text{lat}} = 1$. Thus, the momentum-space representation of the Hamiltonian in Eq.~(\ref{eq:spHamiltonian}) is
\begin{align}
\begin{aligned}
&\hat{\mathcal{H}}_{l=1}^0=\sum_k\hat{\Psi}(k)^\dagger\hat{H}(k)\hat{\Psi}(k), \\&\hat{\Psi}^{\dagger}(k)=(\hat{a}_{k}^{+\dagger},
\hat{a}_{k}^{-\dagger},\hat{b}_{k}^{+\dagger},\hat{b}_{k}^{-\dagger}),
\end{aligned}
\end{align}
with the Bloch Hamiltonian being
 \begin{align}
\begin{split}
    H(k) &= \begin{pmatrix} 
        0 & Q(k) \\ 
        Q^\dagger(k) & 0 
    \end{pmatrix}, \\
    Q(k) &= \begin{pmatrix} 
        A(k) & B(k) \\ 
        C(k) & A(k) 
    \end{pmatrix}.
\end{split}
\label{eq:Kspaceorigin}
 \end{align}
where
\begin{align}
    \begin{aligned}&A(k)=J_2+J_2^{\prime}e^{-ik},\\&B(k)=J_3+J_3^{\prime}e^{-i(k+2\phi)},\\&C(k)=J_3+J_3^{\prime}e^{-i(k-2\phi)}.\end{aligned}
\end{align}
Diagonalization of Eq.~(\ref{eq:Kspaceorigin}) yields four energy bands,
\begin{align}
    E_{1,4}(k)&=\pm\sqrt{\lambda_+(k)},~E_{2,3}(k)=\pm\sqrt{\lambda_-(k)}, \label{eq:E-bands} 
    \end{align}
    with
    \begin{align}
        \lambda_{\pm}(k)&=\frac{2|A|^2+|B|^2+|C|^2}{2} \nonumber
        \\&\pm\sqrt{\left(\frac{|B|^2-|C|^2}{2}\right)^2+\left|AC^*+A^*B\right|^2}. \label{eq:E-lambda}
\end{align}
There is a midspectrum gap closing when the middle pair of bands touch, i.e., when $E_2(k)=E_3(k)$. In Appendix~\ref{app:H_eff_l1}, we detail the derivation of the analytical expression for the critical angle $\phi_c$, which defines the gap-closing point at the critical momentum $k_*$.
We identify two scenarios, depending on whether $k_*$ is independent on $\phi$ or not. First, for $\phi\in(0,\pi/2)$ and $J_{2}^{2}-J_{3}^{2}\ne J_{2}^{\prime2}-J_{3}^{\prime2}$, $k_*=0$ or $\pi$ and the critical angle reads
\begin{align}
    &\phi_c=
    \begin{cases}
    \displaystyle
    \frac{1}{2}\,\arccos\!\left[
        \frac{(J_2+J_2^{\prime})^2-(J_3^{2}+J_3^{\prime2})}
             {2J_3J_3^{\prime}}
    \right], & k_*=0,\\[6pt]
    \displaystyle
    \frac{1}{2}\,\arccos\!\left[
        \frac{(J_{3}^{2}+J_{3}^{\prime2})-(J_{2}-J_{2}^{\prime})^{2}}
             {2J_{3}J_{3}^{\prime}}
    \right], & k_*=\pi.
    \end{cases}
    \label{Eq:NumGeneral}
\end{align}
which only exists when
\begin{align}
    \begin{aligned}
        &|(J_2+J_2^{\prime})^2-(J_3^2+{J_3^{\prime}}^2)|<2|J_3J_3^{\prime}|,~k_*=0,
        \\
        &|(J_3^2+J_3^{\prime2})-(J_2-J_2^{\prime})^2|<2|J_3J_3^{\prime}|,~k_*=\pi.
    \end{aligned}
    \label{Eq:NumGeneralCondition}
\end{align}
On the other hand, when the above relation for the couplings is not fulfilled and $J_{2}^{2}-J_{3}^{2}=J_{2}^{\prime2}-J_{3}^{\prime2}\ne0$, the critical momentum $k_*= k_*(\phi)$ depends on the angle $\phi$. As we show in Appendix~\ref{app:H_eff_l1}, it takes the form:
\begin{align}
    k_*(\phi)=\pm\arccos\!\left[\frac{J_3J_3'\cos(2\phi)-J_2J_2'}{J_2'^2 - J_3'^2}\right],
\end{align}
with the constraint
\begin{align}
\left|\frac{J_3J_3'\cos(2\phi)-J_2J_2'}{J_2'^2 - J_3'^2}\right|\le 1,
\label{Eq:kstarCons}
\end{align}
This dependence implies that there can exist a continuous interval of $\phi\in(0,\pi/2)$ for which the energy gap closes. For each $\phi$ in that interval, one can find a value of $k_*\in(-\pi,\pi]$ satisfying the constraint (\ref{Eq:kstarCons}) and $\phi=\phi_c.$
\par 
Finally, in the special case of $J_{2}^{2}-J_{3}^{2}=J_{2}^{\prime2}-J_{3}^{\prime2}=0$, there is no solution for the critical momentum within the considered range of angles, and the spectrum remains gapped.

\section{Topological invariant and band inversion}\label{sec:TI}

The system under consideration exhibits chiral symmetry [$C H(k)C^{-1}=-H(k)$], time-reversal symmetry [$U_TH^*(k)U_T^\dagger=H(-k),$] and particle–hole symmetry [$U_PH^*(k)U_P^\dagger=-H(-k),$], with the corresponding operators being defined as
\begin{align}
C=
\begin{pmatrix}
I_2 & 0\\ 
0 & -I_2
\end{pmatrix}, 
U_T = \begin{pmatrix}
\sigma_x & 0\\ 
0 & \sigma_x
\end{pmatrix}, U_P =
\begin{pmatrix}
\sigma_x & 0\\
0 & -\sigma_x
\end{pmatrix},
\label{eq:symOperator}
\end{align}
where $I_2$ and $\sigma_x$ are the $2\times2$ identity and $x$ Pauli matrices, respectively. $U_T$ and $U_P$ fulfill $U_TU_T^*=+1$ and $U_PU_P^*=+1$, and therefore the system falls into the topological class BDI \cite{Schnyder2008,Liu2023}. The topological properties of such 1D systems can be characterized by the winding number $W$~\cite{asboth2016}, a $\mathbb{Z}$ invariant. Assuming the mid-gap in the energy spectrum is open, the corresponding winding number can be calculated as follows,
\begin{align}
    W=\frac{1}{2\pi i}\int_{-\pi}^{\pi}dk\frac{d}{dk}\ln\left[\det Q(k)\right].
    \label{Eq:windingNumerical}
\end{align}
According to the bulk-boundary correspondence, the winding number directly determines the number of topologically protected edge states for each boundary. For a finite system with open boundaries, the number of topological edge states in the central gap is $n_{\mathrm{edge}}=2|W|.$
To obtain an analytical expression for $W$, we introduce a new variable $z=e^{-ik}$ with $k\in(-\pi,\pi]$. As the momentum $k$ sweeps from $-\pi$ to $\pi$, $z$ traces the unit circle with $|z|=1$. We introduce this substitution in Eq.~(\ref{Eq:windingNumerical}), which allows to rewrite the winding number as (see Appendix A for details),
\begin{align}
W
=\frac{1}{2\pi i}\oint_{|z|=1}^{CW}\frac{f^{\prime}(z)}{f(z)}dz,
\label{Eq:AnalyticalWinding}
\end{align}
where $f'(z)\equiv \frac{df(z)}{dz}$ and
\begin{align}
    \begin{aligned}
    &f(z)=\det Q(k)
    \\&
=\begin{pmatrix}J_2+J_2^{\prime}z\end{pmatrix}^2-\begin{pmatrix}J_3+J_3^{\prime}e^{-2i\phi}z\end{pmatrix}\begin{pmatrix}J_3+J_3^{\prime}e^{+2i\phi}z\end{pmatrix}.
    \end{aligned}
\end{align}
\par 
This polynomial substitution allows for an analytical derivation of the winding number, whose possible values are limited by the relative strength of the coupling terms. In Appendix~\ref{app:H_eff_l1}, we study the different scenarios in detail, and summarize the different possibilities here,
\begin{align}
W &\in
\begin{cases}
\{0,-1\}, & \left|J_2^2 - J_3^2\right| > \left|J_2'^2 - J_3'^2\right|,\\
\{-1,-2\}, & \left|J_2^2 - J_3^2\right| < \left|J_2'^2 - J_3'^2\right|,\\
\{-1\}, & \left|J_2^2 - J_3^2\right| = \left|J_2'^2 - J_3'^2\right|,
\end{cases}
\label{Eq:wingding}
\end{align}
which are corroborated by the numerical results obtained directly from Eq.~(\ref{Eq:windingNumerical}). From Eq.~(\ref{Eq:wingding}), we observe that the Creutz ladder experiences different kinds of topological transitions as one modifies the relative angle between sites, depending on the coupling values. By tuning these, one can explore both transitions between trivial and nontrivial phases, as well as among nontrivial phases with different number of edge states.
\begin{figure}[H]
    \centering
    \includegraphics[width=1.0\linewidth]{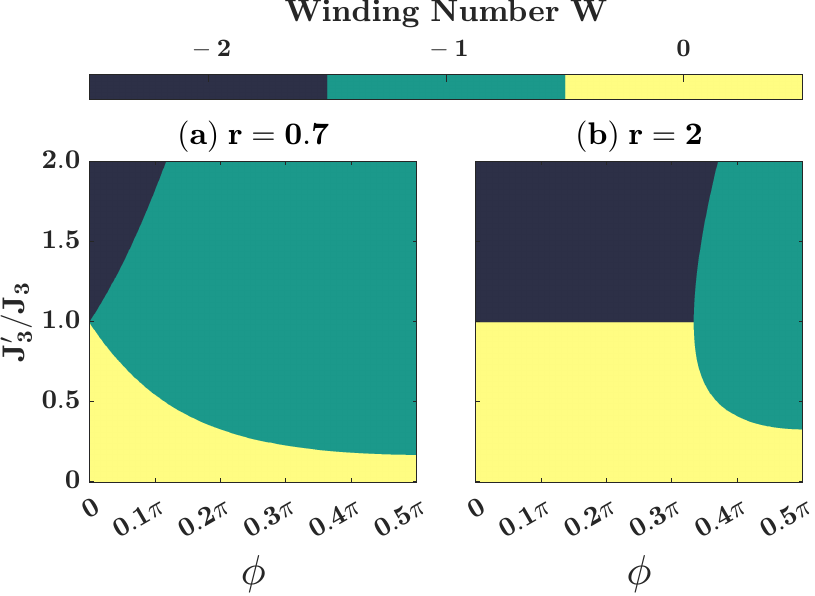}
    \caption{Winding number phase diagram of the system for the fixed coupling ratio $r=J_3/J_2=J'_3/J'_2$ for (a)~$r=0.7$, and (b)~$r=2$, as a function of the angle $\phi$ (horizontal axis) and the coupling ratio $J_3^{\prime}/J_3$ (vertical axis). Regions with different colors correspond to different topological phases according to the value of the winding number indicated on top.}
    \label{fig:2DwindingNumerical}
\end{figure}

In Fig.~\ref{fig:2DwindingNumerical}, we present the phase diagram for the winding number, plotted as a function of the angle $\phi$ and the coupling ratio $J^\prime_3/J_3$. For each panel, we fix a parameter $r=J_3/J_2=J_3^{\prime}/J_2^{\prime}$ and compare the two representative regimes $r=0.7$ and $r=2$. Here, $r$ is a coupling ratio comparing the hopping strength between opposite-circulation and same-circulation states. Although the two values of $r$ produce different diagrams, they have the same trends with respect to $\phi$. For values of $J_3^\prime/J_3<1$, the winding number evolves from $-1$ to $0$ upon sweeping $\phi$, except close to the limit of $J_3^\prime \to 0$. In contrast, for $J_3^\prime/J_3>1$, the winding number changes from $-2$ to $-1$ with increasing $\phi$. It is worth noting that the boundaries between phases in the phase diagram correspond to the gap-closing points of the central two bands, indicating a topological phase transition. A particularly interesting case arises when $J_3^\prime/J_3=1,~r=2$. In this regime,
there exists a critical line joining the $W=0$ and $W=-2$ phases that extends over the whole $\phi \in [0, \phi_c]$ interval. Here, $\phi_c=\pi/3$ is the tri-critical angle where all three phases converge.

The different topological phases can also be understood from the perspective of band inversion~\cite{Liu2020b,Widmann2025}. In our system, the $n$-th band equation for $H(k)$ is
\begin{align}
    H(k)u_n(k)=E_n(k)u_n(k),
\end{align}
with
\begin{align}
    u_n(k)=
\begin{pmatrix}
a_{n,+}(k)\\
a_{n,-}(k)\\
b_{n,+}(k)\\
b_{n,-}(k)
\end{pmatrix}
=
\begin{pmatrix}
a_n(k)\\
b_n(k)
\end{pmatrix}
\end{align}
where $u_n(k)$ is the normalized Bloch eigenvector of the $n$-th band. $a_n(k)$ and $b_n(k)$ denote the two-component amplitudes on the sublattices $a$ and $b$, respectively. Across the topological transition, there exists an exchange of the band character, which can be characterized via the sublattice-exchange operator:
\begin{align}
    \Sigma=
\begin{pmatrix}
0_2&I_2\\
I_2&0_2
\end{pmatrix},
\end{align}
which exchanges the $a$- and $b$-sublattice components of the Bloch eigenvector. Then, the corresponding expectation value of $\Sigma$ is
\begin{align}
    S_n(k)=u_n^\dagger(k)\Sigma u_n(k)=2\operatorname{Re}[a_n^\dagger(k)b_n(k)]. \label{eq:sigma}
\end{align}
For normalized eigenvectors, $|S_n(k)|\le 1$. The value $S_n(k)\approx +1$ indicates that the $a$- and $b$-sublattice components are approximately in phase, while $S_n(k)\approx -1$ indicates that they are approximately out of phase by $\pi$. Therefore, $S_n(k)$ characterizes the relative phase structure between the two sublattice components and serves as the indicator of band inversion, which provides a clear signature of the bulk topology. It is worth to mention that other choices for $\Sigma$ lead to a similar analysis. For instance, one could also choose the operator $\Sigma^\prime=\begin{pmatrix} 0_2 & \sigma_x \\ \sigma_x & 0_2\end{pmatrix}$, which exchanges both the $a$ and $b$ sublattices and the circulations $+$ and $-$.

\begin{figure}[t]
    \centering
    \includegraphics[width=1.0\linewidth]{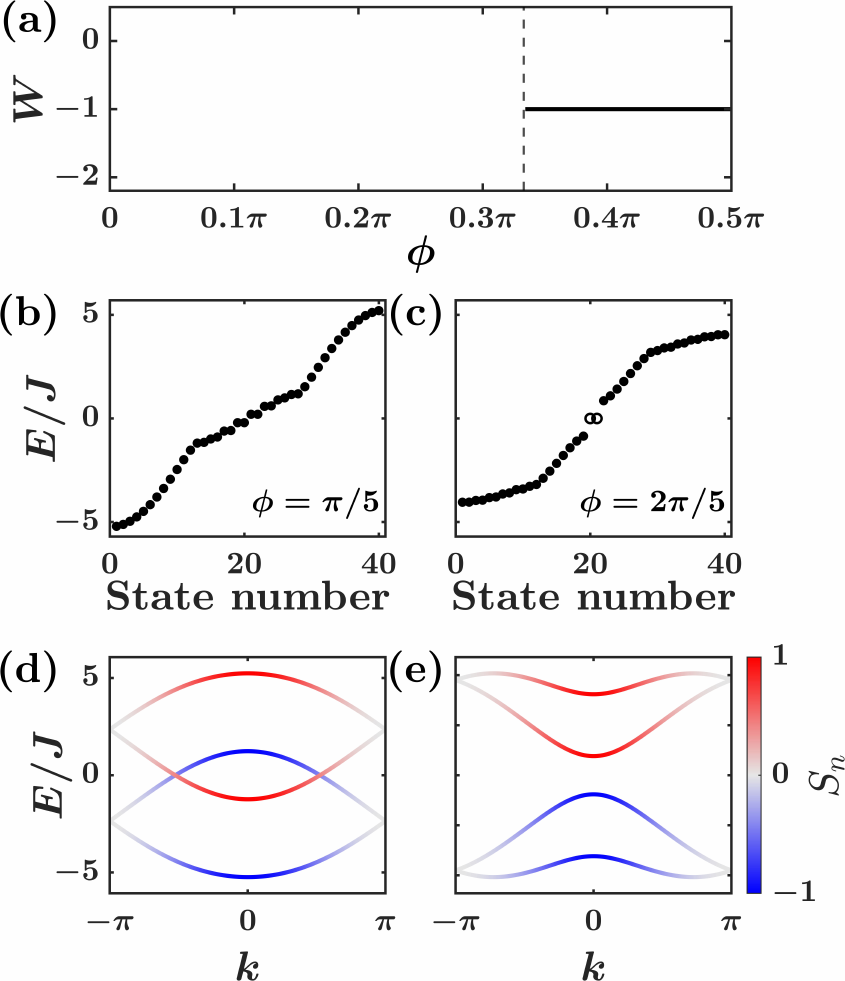}
    \caption{(a) Winding number $W$, calculated from Eq.~(\ref{Eq:windingNumerical}), as a function of the angle $\phi$. The vertical dashed line marks the critical angle $\phi_c=\pi/3$ above which $W$ is well-defined. 
    Energy spectrum for $N_c = 10$ unit cells with (b) $\phi=\pi/5$, showing a gapless case without topological edge states and (c) $\phi=2\pi/5$, where $W=-1$ and two topological edge states exist (hollow symbols). (d)--(e) Band structures for periodic boundary conditions, with the colorbar representing the signature of band inversion $S_n(k)$, for the same angles $\phi$ as (b) and (c), respectively.
    Hopping parameter values: $J_2=J'_2=J,~J_3=J'_3=2J$.
    }
    \label{fig:gaplessTotwo}
\end{figure}

\section{Numerical results}\label{sec:boson}

In this section, we discuss the results for the tight-binding Hamiltonian of the system. In the following, the discussion is divided into subsections classified by the number of topological edge states that the system can display, and the type of topological transition that occurs, or lack thereof. Within each subsection, we fix the couplings values and vary the angle $\phi$ to track the changes in the topological properties of the system. Our discussion covers the three representative cases summarized in Eq.~(\ref{Eq:wingding}). In subsection A, the emergence of edge states; in subsection B, the transition between different nontrivial phases; and in subsection C, the long distance limit, for which no transition occurs. 
The following results are obtained numerically from Eq.~(\ref{Eq:windingNumerical}), and show complete agreement with the analytical predictions. We further show the corresponding energy spectrum of a finite open chain obtained by exact diagonalization, which confirms the bulk–edge correspondence.

\subsection{Emergence of edge states}\label{sec:n_edge_0_2}

We start by treating the case in which the system transitions from 
zero to two edge states. There are two possible scenarios. First, for $\left|J_2^2 - J_3^2\right| = \left|J_2'^2 - J_3'^2\right|$, the system is gapless within a certain angle interval, and thus the winding number is not well defined, indicating an absence of edge states. As shown in Fig.~\ref{fig:gaplessTotwo}(a), for $J_2=J'_2=J$ and $~J_3=J'_3=2J$ this interval is $0<\phi<\pi/3$. Once $\phi$ exceeds the critical angle $\phi_c=\pi/3$, the system becomes fully gapped with winding number $W=-1$. 
\begin{figure}[t]
    \centering
    \includegraphics[width=1.0\linewidth]{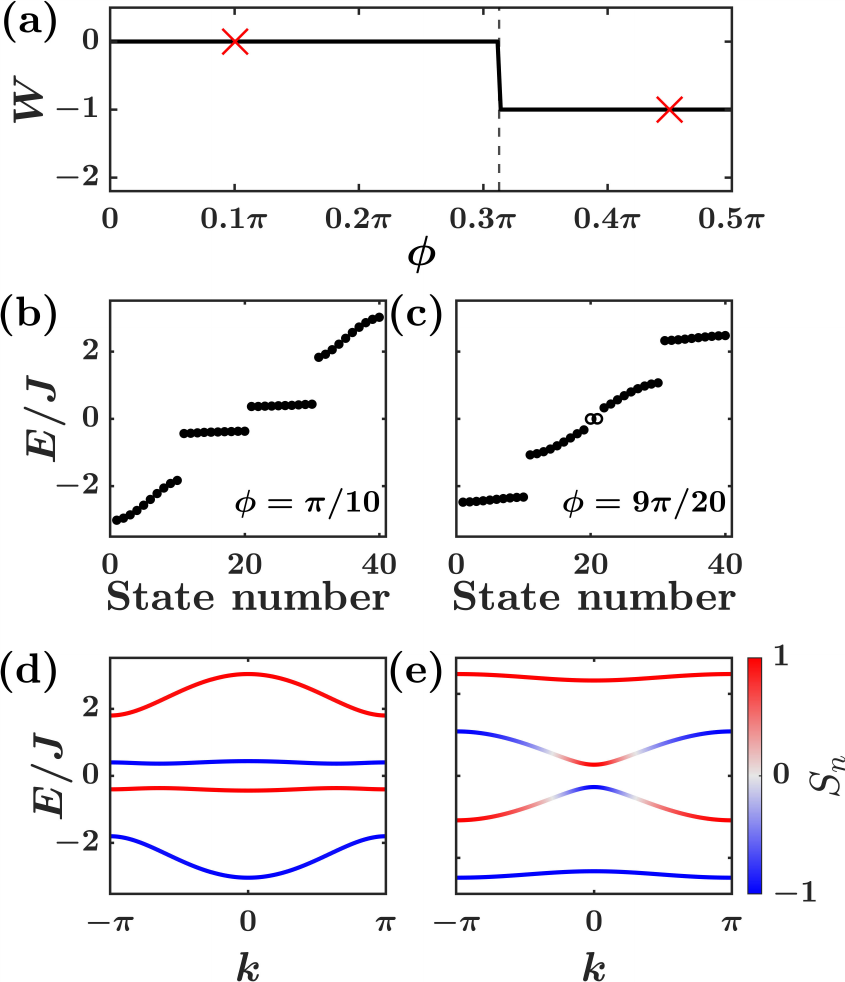}
    \caption{(a) Winding number $W$ as a function of the angle $\phi$. The vertical dashed line marks the critical angle $\phi_c\approx 5\pi/16$ where $W$ changes and the red crosses mark $\phi=\pi/10$ and $\phi=9\pi/20$. 
    Energy spectrum for $N_c = 10$ unit cells with (b) $\phi=\pi/10$, showing a case $W=0$ without topological edge states and (c) $\phi=9\pi/20$, where $W=-1$ and two topological edge states emerge~(hollow symbols).
    (d)--(e) Band structures for periodic boundary conditions, with the colorbar representing the signature of band inversion $S_n(k)$, for the same angles $\phi$ as (b) and (c), respectively.
    Hopping parameter values: $J_2=J,~J'_2=0.3J,~J_3=1.4J,~J'_3=0.4J$.}
    \label{fig:WzerotoOne}
\end{figure}
Fig.~\ref{fig:gaplessTotwo}(b) shows the energy spectrum for $\phi=\pi/5$ (in the gapless regime), where all eigenstates are embedded in the bulk bands and no topological edge state exists. In contrast, Fig.~\ref{fig:gaplessTotwo}(c) shows the spectrum for $\phi=2\pi/5$, where the gap is opened and two topological edge states appear. These differences in the energy spectrum are consistent with values of the winding number in Fig.~\ref{fig:gaplessTotwo}(a). Figs.~\ref{fig:gaplessTotwo}(d) and (e) show the bulk band structure for the same angle $\phi$ as (b) and (c), respectively, where the bands are colored by $S_n(k)$ as defined in (\ref{eq:sigma}). Red and blue correspond to the positive and negative $S_n(k)$ regions, respectively, implying that the $a$- and $b$-sublattice components of the Bloch eigenvector are mainly in phase for the former and in anti-phase for the latter. The two central bands in Fig.~\ref{fig:gaplessTotwo}(d) exchange the value of $S_n(k)$ around $k=0$ when the gap opens at $\phi_c=\pi/3$, thus showcasing band inversion. This inversion accompanies the change in winding number and the emergence of the topological edge states shown in Figs.~3(b) and 3(c).

\begin{figure}[t]
    \centering
    \includegraphics[width=1.0\linewidth]{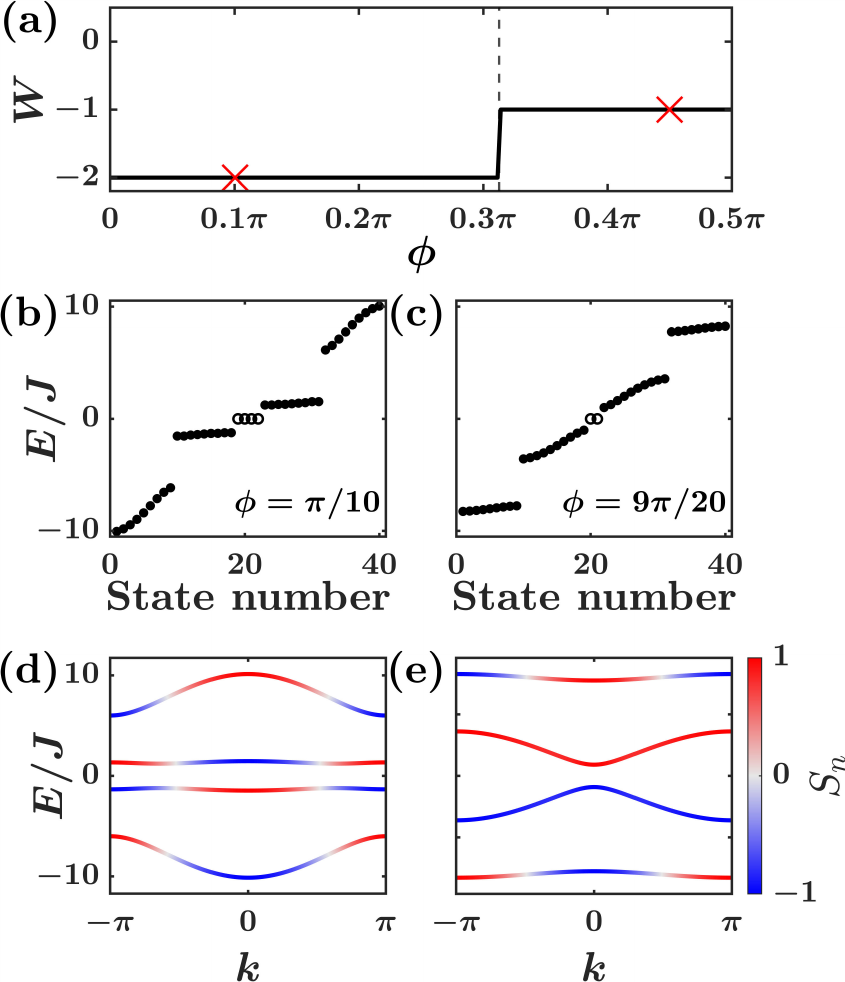}
     \caption{(a) Winding number $W$ as a function of the angle $\phi$. The vertical dashed line marks the critical angle $\phi_c\approx 5\pi/16$ where $W$ changes and the red crosses mark $\phi=\pi/10$ and $\phi=9\pi/20$. 
     Energy spectrum for $N_c = 10$ unit cells with (b) $\phi=\pi/10$, showing a case with $W=-2$ and four topological edge states~(hollow symbols) and (c) $\phi=9\pi/20$, with $W=-1$ and two topological edge states~(hollow symbols).
     (d)--(e) Band structures for periodic boundary conditions, with the colorbar representing the signature of band inversion $S_n(k)$, for the same angles $\phi$ as (b) and (c), respectively.
     Hopping parameter values: $J_2=J,~J'_2=\frac{10}{3}J,~J_3=\frac{4}{3}J,~J'_3=\frac{14}{3}J$.
     }
    \label{fig:WindingTwotoOne}
\end{figure}

Secondly, we present the case in which the system undergoes a topological transition from $W=0$ to $W=-1$, which occurs when $\left|J_2^2 - J_3^2\right| > \left|J_2'^2 - J_3'^2\right|$.  Fig.~\ref{fig:WzerotoOne}(a) shows the winding number as a function of the angle for $J_2=J,~J'_2=0.3J,~J_3=1.4J,~J'_3=0.4J$, with the critical angle in this case being $\phi_c\approx 5\pi/16$.
Since the condition $J_{2}^{2}-J_{3}^{2}\neq J_{2}^{\prime2}-J_{3}^{\prime2}$ is satisfied in this case, we can use the analytical prediction for the critical angle $\phi_c$ given by Eq.~(\ref{Eq:NumGeneral}). From Eq.~\ref{Eq:NumGeneralCondition}, it follows that the gap-closing point appears only at $k=0$, leading to $\phi_c=\arccos\sqrt{\frac{69}{224}}\approx5\pi/16$, which is in full agreement with the numerical calculation shown in Fig.~\ref{fig:WzerotoOne}(a). Fig.~\ref{fig:WzerotoOne}(b) presents the energy spectrum at $\phi=\pi/10$, clearly showing an open gap around $E = 0$ and the absence of topological edge states, consistent with $W=0$. 
For the case corresponding to $\phi=9\pi/20$ shown in Fig.~\ref{fig:WzerotoOne}(c), the energy spectrum reveals two topological edge states (hollow symbols), fully consistent with the nontrivial winding number $W=-1$. Figs.~\ref{fig:WzerotoOne}(d) and (e) show band inversion in the central gap for the same angles as (b) and (c), respectively, in line with the change in the winding number.

\subsection{Transition between different nontrivial phases}\label{sec:n_edge_2_4}

We now jump to the case where the transition joins topological phases with a different number of edge states. Here, we focus on the case satisfying $\left|J_2^2 - J_3^2\right| < \left|J_2'^2 - J_3'^2\right|$. Specifically, we choose $J_2=J,~J'_2=\frac{10}{3}J,~J_3=\frac{4}{3}J,~J'_3=\frac{14}{3}J$.
From Eqs.~(\ref{Eq:NumGeneral}) and (\ref{Eq:NumGeneralCondition}) we know that the analytical critical angle $\phi_c$ has the same value as the previous scenario, $\phi_c=\arccos\sqrt{\frac{69}{224}}$, with the gap closing at $k=0$. Fig.~\ref{fig:WindingTwotoOne}(a) shows the winding number as a function of angle $\phi$ for this case, which changes from $-2$ to $-1$ as $\phi$ varies, with the system remaining nontrivial except for the gap closing point. Fig.~\ref{fig:WindingTwotoOne}(b) presents the energy spectrum for the angle $\phi=\pi/10$, which clearly shows that there are four topological edge states in the gap, consistent with the winding number $W=-2$. In contrast, Fig.~\ref{fig:WindingTwotoOne}(c) shows the spectrum at $\phi=9\pi/20$, exhibiting only two edge states in agreement with $W=-1$. As in the cases discussed above, as displayed in Figs.~\ref{fig:WindingTwotoOne}(d) and (e), there is an exchange of the value of $S_n(k)$ in the central gap, consistent with the decrease of $|W|$.

\begin{figure}[H]
    \centering
    \includegraphics[width=1.0\linewidth]{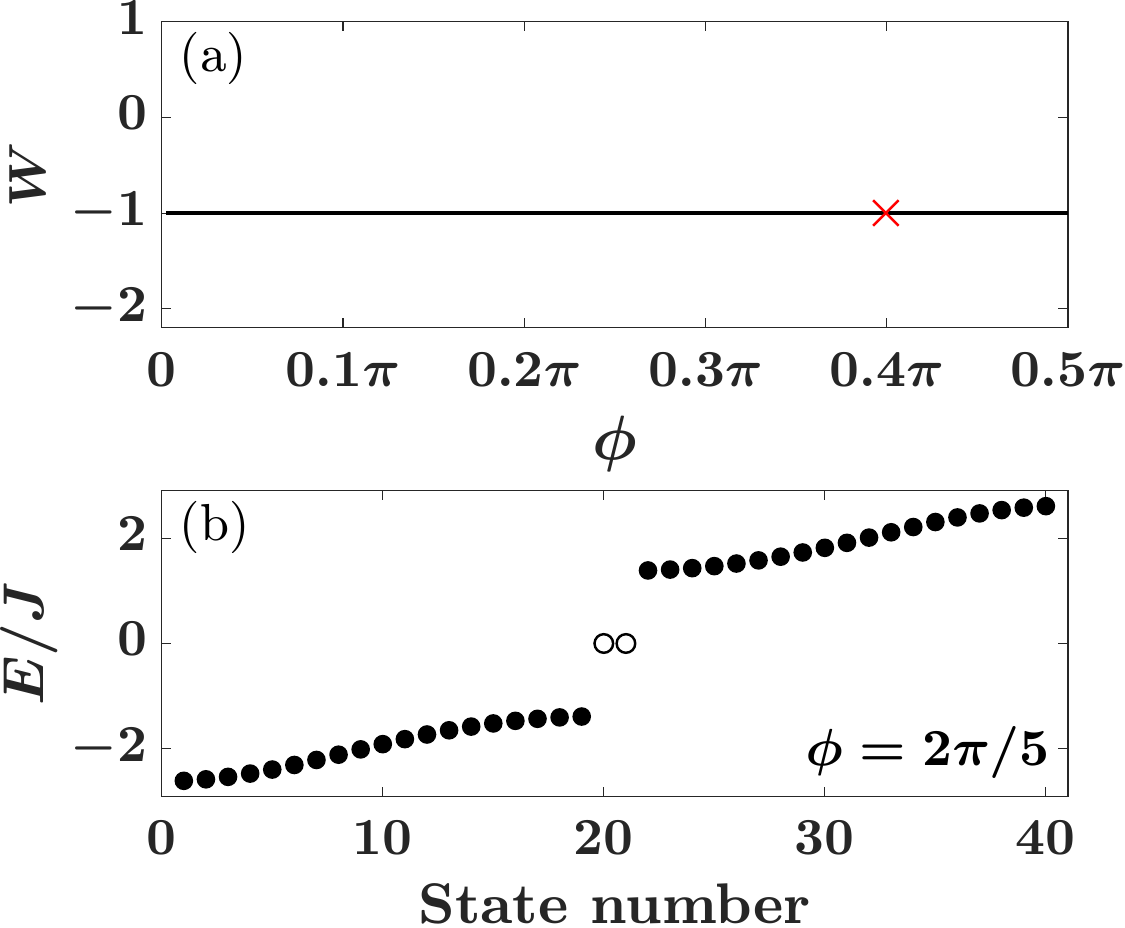}
     \caption{(a) Winding number $W$ as a function of the angle $\phi$. The red cross mark $\phi=2\pi/5$. 
     Energy spectrum for (b) $\phi=2\pi/5$, where $W=-1$ and two topological edge states exist within the gap. Hopping parameter values: $J_2=J'_2=J_3=J'_3=J$ and $N_c = 10$ unit cells.
     }
    \label{fig:WindingOne}
\end{figure}
\subsection{Long distance limit} \label{sec:longdistance}

Let us now consider our system in the long-distance limit, where both distances $d$ and $d'$ are sufficiently large and all the couplings become equal, i.e., $J_2=J'_2=J_3=J'_3=J$. Fig.~\ref{fig:WindingOne}(a) shows the winding number as a function of the angle $\phi$, where one observes that it remains constant at $W=-1$ for the entire range of angles. This is in agreement with the analytical prediction in Sec.~\ref{sec:TI} that there is no critical angle in this case. As shown in Fig.~\ref{fig:WindingOne}(b), the energy spectrum for $\phi=2\pi/5$ exhibits two topological edge states, as it does for all values of the angle in the considered range. Finally, we note that no band inversion occurs in this case. Since there is no change in the winding number, $S_n(k)$ likewise keeps a constant parity for all angles.

\section{Implementation proposal}\label{sec:waveguides}

A platform that naturally allows for the exploration of the described model is an array of coupled cylindrical optical waveguides \cite{Jorg2020,Jiang2023,Wang2024,Viedma2024b}, for which the set of OAM modes forms an eigenbasis. The well-known quantum-optical analogy \cite{Longhi2009} establishes that the description of the temporal evolution of a quantum system is formally equivalent to that of light propagation in an analogous photonic system. Hence, optical waveguides constitute a remarkably flexible platform for the implementation and study of noninteracting bosonic systems, as they allow for precise control over their coupling parameters and for a simple introduction of AGFs. This has led to a wide range of works focused on various TIs and flat-band systems \cite{Leykam2018b,Vicencio2021,Caceres-Aravena2022,Schulz2022}.

\begin{figure}[t]
    \centering
    \includegraphics[width=\linewidth]{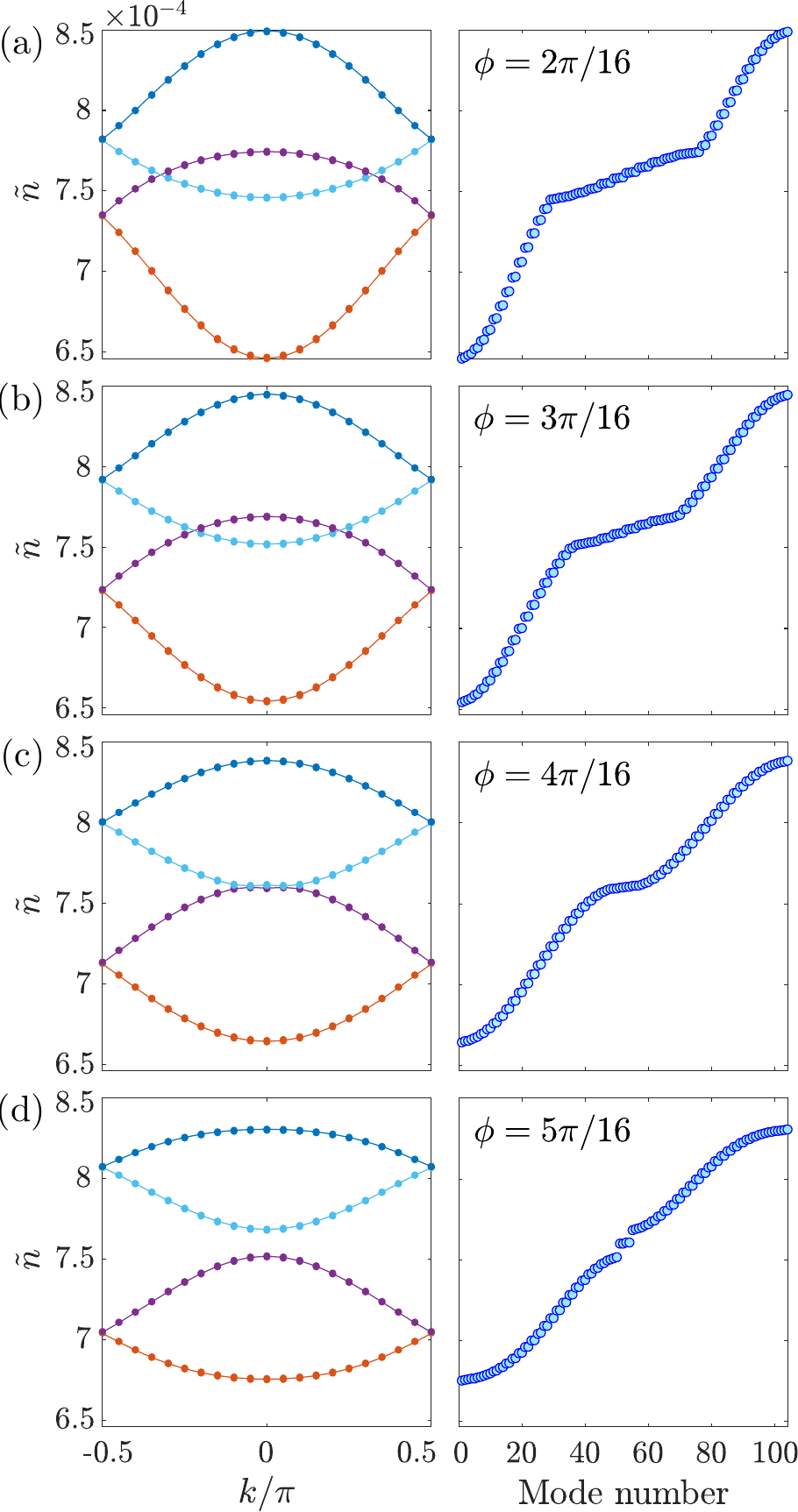}
    \caption{Band structure of effective mode indices $\tilde{n} = n_{\text{eff}} - n_0$ for the zig-zag photonic chain for constant waveguide distances $d = d'=\SI{12}{\micro\meter}$, for the periodic system (left column) and the finite system with $N_c=13$ unit cells (right column). Increasing values of $\phi$ are considered in each row, according to (a) $\phi=2\pi/16$, (b) $\phi=3\pi/16$, (c) $\phi=4\pi/16$ and (d) $\phi=5\pi/16$, sweeping across the gap-opening point.}
    \label{fig:comsol_d60}
\end{figure}

We build finite-element simulations for the model in Fig.~\ref{fig:l1model}(a) using the commercial solver COMSOL Multiphysics. We consider sets of waveguides guiding up to both circulations of $l=1$ modes, each with a radius of $R = \SI{4}{\micro\meter}$, a cladding refractive index of $n_0 = 1.48$ and an index contrast of $\Delta n = 3\cdot 10^{-4}$. Then, we choose different configurations to explore the various coupling regimes, both for periodic and finite chains. We start by building a periodic zig-zag chain for a constant distance between waveguides of $d = \SI{12}{\micro\meter}$, corresponding to a coupling ratio of $J_2 = J_2^\prime \equiv J$ and $J_3 = J_3^\prime \simeq 1.44 J$ \cite{Viedma2024b}, which falls under the umbrella of Sec.~\ref{sec:n_edge_0_2}. In the left column of Fig.~\ref{fig:comsol_d60}, we plot the band structure of the system as we sweep the angle $\phi$ across the critical angle $\phi_c \simeq \pi/4$, where the opening of the central gap is readily observed. Afterwards, we consider a finite chain of $N=13$ unit cells and simulate the spectrum for the same angles in the right column of Fig.~\ref{fig:comsol_d60}, which clearly reveals the appearance of two edge states for $\phi>\phi_c$ in the central gap. Note that the points are doubled as the eigensolver yields the eigenmodes for the two orthogonal field polarizations that can be guided by the structure, and thus four points appear in the gap of Fig.~\ref{fig:comsol_d60}(d). 

\begin{figure}[t]
    \centering
    \includegraphics[width=\linewidth]{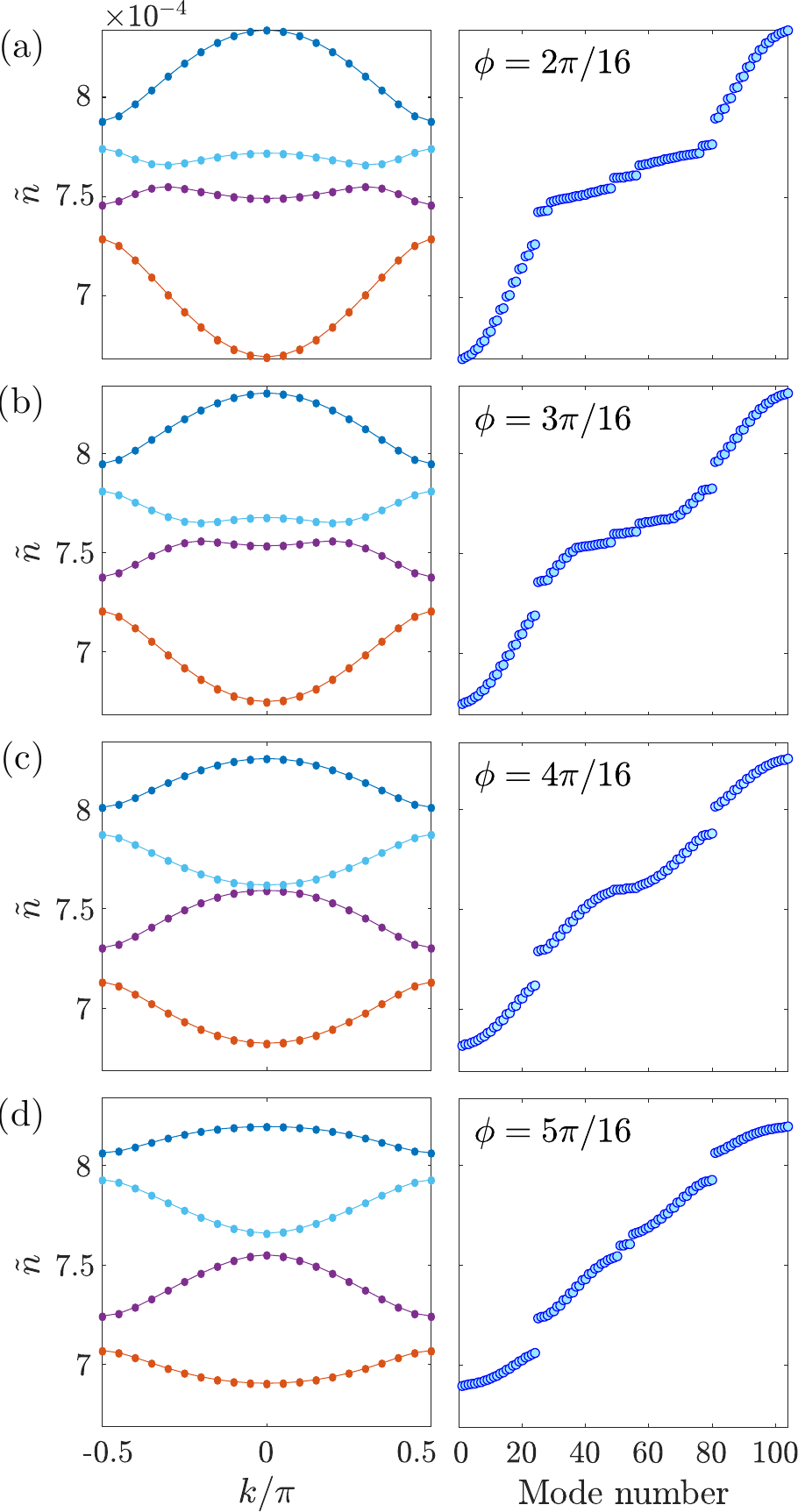}
    \caption{Band structure of effective mode indices $\tilde{n} = n_{\text{eff}} - n_0$ for the zig-zag photonic chain with alternating waveguide distances $d = \SI{13}{\micro\meter}$ and $d' = \SI{12}{\micro\meter}$, for the periodic system (left column) and the finite system with $N_c=13$ unit cells (right column). Increasing values of $\phi$ are considered in each row, according to (a) $\phi=2\pi/16$, (b) $\phi=3\pi/16$, (c) $\phi=4\pi/16$ and (d) $\phi=5\pi/16$, sweeping across the topological transition.}
    \label{fig:comsol_d65_60}
\end{figure}

We now go back to the periodic chain, but this time considering alternating distances, $d = \SI{13}{\micro\meter}$ and $d' =\SI{12}{\micro\meter}$, which corresponds to the following coupling ratios: $J_2 = J$, $J_3 \simeq 1.41 J$, $J_{2}'\simeq 1.57J$ and $J_{3}' \simeq 2.26 J$ \cite{Viedma2024b}. This scenario corresponds to Sec.~\ref{sec:n_edge_2_4}, and therefore one expects a topological transition between two nontrivial phases with different winding numbers. One should note that reversing the distances $d\leftrightarrow d'$ gives access to the second case explored in Sec.~\ref{sec:n_edge_0_2} instead.
First, the band structure of the periodic system in the left part of Fig.~\ref{fig:comsol_d65_60} reveals that there exists a gap closing point again around $\phi_c \simeq \pi/4$, which this time separates two gapped regimes. When considering the finite system in Fig.~\ref{fig:comsol_d65_60}, it is clear that at that point we jump from four to two edge states for each polarization. 


\begin{figure*}[t]
    \centering
    \includegraphics[width=0.95\linewidth]{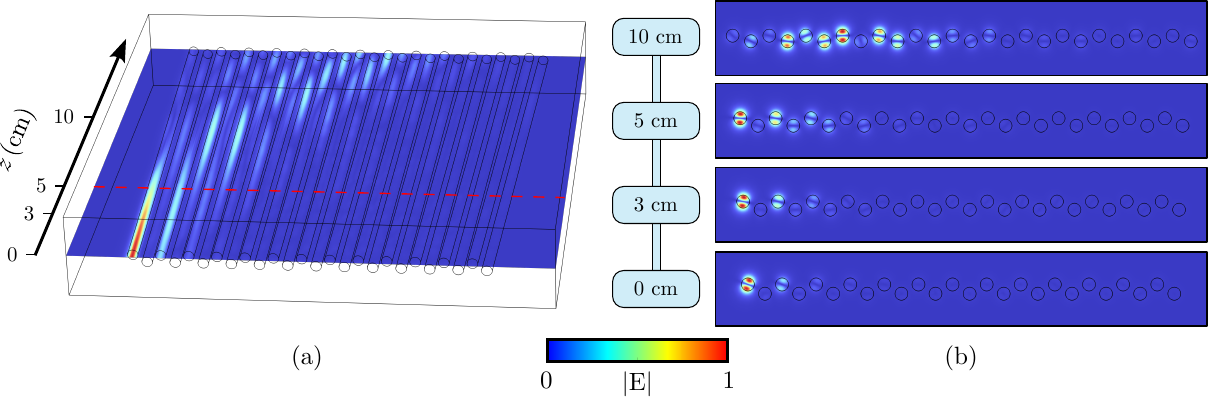}
    \caption{Electric field propagation in a zig-zag chain where the angle between waveguides changes along the propagation direction from $\phi_0 = 5\pi/16$ to $\phi_F = 2\pi/16$. (a) Electric field in a cross-section of the device, with the point where the topological transition occurs ($z\approx \SI{5}{cm}$) marked with a red dashed line. (b) Snapshots at different propagation lengths before and after the topological transition. Both visualizations clearly show the localization being lost for $z>\SI{5}{cm}$.}
    \label{fig:propagation}
\end{figure*}

Finally, we consider light propagation in the waveguide system. Beyond the calculation of the eigenmode spectrum, a way to observe the existence of the topological transition is by gradually changing the angle between waveguides along the propagation direction. We consider the case where all distances are equal $d = \SI{12}{\micro\meter}$, and start with an angle that is larger that the critical angle $\phi_0 > \phi_c$. Then, we stretch the chain along $z$, reducing this angle while keeping the distances constant, so that at the end facet the angle of the chain is $\phi_F < \phi_c$. In such a situation, an initial excitation on an edge mode remains localized during propagation up until the topological transition, where the mode disappears from the spectrum and light disperses into the bulk. We observe precisely this behavior in Fig.~\ref{fig:propagation}, where light propagation is simulated for an initial angle of $\phi_0 = 5\pi/16$ and a final angle of $\phi_F = 2\pi/16$, for a device of length $L=\SI{15}{cm}$ where the transition point occurs around $z\approx \SI{5}{cm}$. A qualitatively similar result occurs if only the edgemost waveguide is excited and not the exact edge mode with the correct amplitudes and phases in different waveguides. This is a natural effect of the edge state having the largest amplitude in the final waveguide. In this second situation, the small fraction of light that does not overlap with the edge mode disperses into the bulk from the start, while the rest follows the same behavior as Fig.~\ref{fig:propagation}.

Since the photonic device is entirely reciprocal, one could in theory reverse the device and use it to obtain a precise localization of light from an initially delocalized input, which would also prove the topological transition. However, this would require engineering the input beam in the device to heavily overlap with the edge mode at the point of transition, as otherwise only a low fraction of light would reach the mode and would make observation challenging. For the purpose of precisely exciting topological modes in waveguide systems, with the simplest input beams possible, other methods have already been proposed \cite{Viedma2022,Viedma2023,Liu2024,Real2024,Liu2025}.

\section{Conclusions}\label{sec:conclusions}

We have investigated a 1D staggered ladder structure, and worked within the subspace of OAM states with charge $l=1$.
The system can be mapped into a Creutz ladder where the OAM circulations behave as a synthetic dimension, and with the phase of the cross-hopping terms being controlled by the interleg angle of the ladder.
We have calculated the winding number $W$ to study the topological properties of the system. By deriving its analytical form, we reveal that its possible values are determined by the relative strengths of the hopping amplitudes.
Then, we have proved that the topological phase of the system may be changed, for the same hopping values, simply by tuning the interleg angle $\phi$. Hence, by modifying this angle, a topological phase transition that changes the number of edge states can be induced.
For equally separated sites, for which the gap is closed at low angles and $W$ is ill-defined, we instead observe a gap opening with a change in the number of edge states instead of a transition.
In contrast, in the long-distance limit, the system remains in the topological phase with $W=-1$ for the whole range $\phi\in(0,\pi/2)$, and no critical angle of transition exists. The analytical predictions have been further confirmed by numerical calculations. Additionally, the changes in winding number have been related to the emergence of band inversion in the central energy gap. For each topological transition, the expectation value of the sublattice-exchange operator changes parity for each band, and therefore we conclude that band inversion can be used as a topological marker.

To implement the discussed system experimentally, we have proposed a chain of cylindrical optical waveguides that naturally host the required OAM modes. Using commercial simulation software, we have built finite-element simulations for different waveguide couplings, allowing for the exploration of the two mentioned transitions both for finite and periodic systems, showing complete agreement with the theoretical predictions. Moreover, we have shown that the topological transitions can be observed via light propagation in such waveguide systems. 
In a device where the waveguide angle is slowly modified along the propagation direction, a beam initially localized in a topological edge state is shown to delocalize as the transition occurs.
\par
In recent years, the usage of OAM states has become available in several experimental platforms. In ultracold atom systems, OAM states in ring potentials can be generated through a temperature quench~\cite{Corman2014}, by rotating a weak link around the ring~\cite{Wright2013}, or via a two-photon Raman process~\cite{Andersen2006}. In photonic lattices, light beams carrying OAM can be injected into arrays of cylindrical waveguides~\cite{Jorg2020}. Within that context, our work introduces a new way to achieve topological control by tuning the lattice angle, which could be explored with the above schemes. 
This idea could also be extended to other models, such as diamond chain structures~\cite{Pelegri2019}, trimer lattices~\cite{Martinez2019} or the SSH4 model~\cite{xie2019}. 
Furthermore, this concept can be generalized to systems with strong interactions. Long-range interactions have been realized in ultracold atoms~\cite{Annette2025,Baier2016} and photonic systems~\cite{roushan2017,morvan2022}. Recent work~\cite{Salerno2020} has proposed an effective lattice model describing the center-of-mass motion for bound states, offering a theoretical framework for two-body systems where our work could find application.

\section{Acknowledgments}

The authors acknowledge Ayaka Usui for useful discussions.
Y.Z, V.A. and D.V. acknowledge financial support from the Spanish Ministerio de Ciencia e Innovación (MCIN) (MCIN/AEI/10.13039/501100011033, contract No.~PID2020-118153GBI00 and PID2024-160393NB-I00) and Generalitat de Catalunya (Contract No.~SGR2021-00138). Y.Z. acknowledges the Scholarship under the Grant N.202306890031.
D.V. acknowledges funding from MCIN (MCIN/AEI/10.13039/501100011033, contract No.~PID2023-149988NB-C21) and funding from European Union NextGenerationEU (PRTR-C17.I1).
A.M.M. and R.G.D. developed their work within the scope of Portuguese Institute for Nanostructures, Nanomodelling and Nanofabrication (i3N) Projects No.~UIDB/50025/2020, No.~UIDP/50025/2020, and No.~LA/P/0037/2020, financed by national funds through the Funda\c{c}\~{a}o para a Ci\^{e}ncia e Tecnologia (FCT) and the Minist\'{e}rio da Educa\c{c}\~{a}o e Ci\^{e}ncia (MEC) of Portugal. A.M.M. acknowledges financial support from i3N through the work Contract No.~CDL-CTTRI-91-SGRH/2024.

\appendix
\section{Derivation of the analytical form of the winding number and the critical angle for the gap-closing point} \label{app:H_eff_l1}
Eq.~\ref{Eq:windingNumerical} for the winding number can be rewritten as
\begin{align}
    W=\frac{1}{2\pi i}\int_{-\pi}^{\pi}dk\frac{d}{dk}\ln f(k),\quad f(k)=\det Q(k).
    \label{ap1}
\end{align}
Following the derivation in the main text, we set $z = e^{-ik}$ with $k \in [-\pi, \pi]$. As $k$ varies from $-\pi$ to $\pi$, the complex variable $z$ traces the unit circle $|z| = 1$ in the complex plane. The derivative of $z$ with respect to $k$ is $\frac{dz}{dk} = -i e^{-ik} = -i z$, which leads to
\begin{align}
    dz = -i z\, dk, \qquad dk = i\, \frac{dz}{z}.
\end{align}
For an analytic function $f(k)$, the derivative with respect to $k$ can be expressed in terms of $z$ as
\begin{align}
\begin{aligned}
    \frac{d}{dk} \ln f(k)
    &= \frac{dz}{dk}\, \frac{d}{dz} \ln f(z)
    = (-i e^{-ik})\, \frac{1}{f(z)} \frac{df}{dz}
    \\&
    = -i z\, \frac{f'(z)}{f(z)}.
    \end{aligned}
\end{align}
This transformation allows us to rewrite $k$-space integrals over the Brillouin zone as contour integrals along the unit circle in the complex $z$-plane. Then, the winding number in Eq.~\ref{ap1} can be rewritten as,
\begin{align}
W=\frac{1}{2\pi i}\oint_{|z|=1}^{CW}\frac{f^{\prime}(z)}{f(z)}dz.
\label{A5}
\end{align}
where $CW$ indicates a clockwise direction of the unit circle.
\par 
We now specify the analytic function as
\begin{align}
    f(z)=\alpha_0+\alpha_1(\phi)z+\alpha_2z^2,
    \label{A6}
\end{align}
with the coefficients
\begin{align}
\begin{aligned}
    \alpha_0 &= J_2^2 - J_3^2 ,\\[4pt]
    \alpha_1(\phi) &= 2\left(J_2 J_2' - J_3 J_3' \cos 2\phi \right),\\[4pt]
    \alpha_2 &= J_2'^2 - J_3'^2 .
\end{aligned}
\label{eq:defalpha}
\end{align}

In the following, we will restrict our analysis to $\phi\in(0,\pi/2).$

When $\alpha_{2}\neq0$, the function $f(z)$ can be factorized as
\begin{align}
\begin{aligned}
    f(z)&=\alpha_2\left(z-z_1\right)(z-z_2).
\end{aligned}
\end{align}
where the roots are
\begin{align}
    z_{1,2}=\frac{-\alpha_1\pm\sqrt{\alpha_1^2-4\alpha_0\alpha_2}}{2\alpha_2}.
\end{align}
Thus,
\begin{align}
\begin{aligned}
    \frac{f^{\prime}(z)}{f(z)}=\frac{d}{dz}\ln\left[\alpha_{2}(z-z_{1})(z-z_{2})\right]=\frac{1}{z-z_{1}}+\frac{1}{z-z_{2}},
    \end{aligned}
\end{align}
and substituting this expression into Eq.~\ref{A5},
\begin{align}
\begin{aligned}
    W=&\frac{1}{2\pi i}\oint_{|z|=1}^{CW}\left[\frac{1}{z-z_{1}}+\frac{1}{z-z_{2}}\right]dz\\=&\begin{cases}-2,&|z_1|<1,|z_2|<1
    \\-1,&\min(|z_1|,|z_2|)<1<\max(|z_1|,|z_2|)
    \\0,&|z_1|>1,|z_2|>1
    \end{cases}
    \end{aligned}
\end{align}
Here, the value of $W$ is determined by the number of zeros of $f(z)$ enclosed within the unit circle.
\par 
If $\alpha_{2}=0$, the function $f(z)$ reduces to a linear form,
\begin{align}
    f(z)=\alpha_0+\alpha_1(\phi)z,
\end{align}
and the winding number becomes
\begin{align}
\begin{aligned}
    W=&\frac{1}{2\pi i}\oint_{|z|=1}^{CW}\frac{\alpha_1(\phi)}{\alpha_0+\alpha_1(\phi)z}dz&\\
    =&\begin{cases}-1,&|\alpha_0|<|\alpha_1(\phi)|\\0,&|\alpha_0|>|\alpha_1(\phi)|\end{cases}
    \end{aligned}
\end{align}
In summary, the winding number can be generally expressed as 
\begin{align}
W &\in
\begin{cases}
\{0,-1\}, & \left|J_2^2 - J_3^2\right| > \left|J_2'^2 - J_3'^2\right|,\\
\{-1,-2\}, & \left|J_2^2 - J_3^2\right| < \left|J_2'^2 - J_3'^2\right|,\\
\{-1\}, & \left|J_2^2 - J_3^2\right| = \left|J_2'^2 - J_3'^2\right|.
\end{cases}
\end{align}
\par 
The winding number is well-defined only when the bulk energy gap remains open. When the gap closes, the winding number becomes ill-defined. In our model, the gap closes if there exists a momentum $k$ such that $\lambda_-(k)=0$ or, equivalently, when $f(z)$ has a zero value on the unit circle $|z|=1$. Thus, the gap-closing condition can be derived from
\begin{align}
\begin{aligned}
    &(\alpha_2-\alpha_0)\sin k=0,\\~&(\alpha_2+\alpha_0)\cos k +\alpha_1(\phi)=0.
    \end{aligned}
    \label{A14}
\end{align}
When $\alpha_2\ne \alpha_0$, the first equation in Eq.~\ref{A14} yields $\sin k=0$, implying $k=0$ or $\pi$, The second equation then becomes
\begin{align}
\begin{aligned}
    &\alpha_0+\alpha_1(\phi)+\alpha_2=0\quad(k=0),\\
    &\alpha_0-\alpha_1(\phi)+\alpha_2=0\quad(k=\pi).
\end{aligned}
\end{align}
Solving for the critical angle $\phi_c$, one finds
\begin{align}
    &\phi_c=
    \begin{cases}
    \displaystyle
    \frac{1}{2}\,\arccos\!\left[
        \frac{(J_2+J_2^{\prime})^2-(J_3^{2}+J_3^{\prime2})}
             {2J_3J_3^{\prime}}
    \right], & k=0,\\[6pt]
    \displaystyle
    \frac{1}{2}\,\arccos\!\left[
        \frac{(J_{3}^{2}+J_{3}^{\prime2})-(J_{2}-J_{2}^{\prime})^{2}}
             {2J_{3}J_{3}^{\prime}}
    \right], & k=\pi.
    \end{cases}
\end{align}
with the condition
\begin{align}
    \begin{aligned}
        &|(J_2+J_2^{\prime})^2-(J_3^2+{J_3^{\prime}}^2)|< 2|J_3J_3^{\prime}|,~k=0,
        \\
        &|(J_3^2+J_3^{\prime2})-(J_2-J_2^{\prime})^2|<2|J_3J_3^{\prime}|,~k=\pi.
    \end{aligned}
\end{align}
In the special case where $\alpha_2= \alpha_0\ne0$, the first equation in Eq.~\ref{A14} can always be satisfied. In this case, the gap-closing condition simplifies to
\begin{align}
    \cos k=-\frac{\alpha_1(\phi)}{2\alpha_2},
    \label{eq:a2a0ne0}
\end{align}
and the corresponding critical angle reads
\begin{align}
    \phi_c = \frac{1}{2} \arccos\left[ \frac{ (J_{2}^{\prime2}-J_{3}^{\prime2})\cos k + J_2 J_2^{\prime} }{ J_3 J_3^{\prime} } \right],
\end{align}
with the constraint
\begin{align}
    |\alpha_2\cos k+J_2J_2^\prime|<|J_3J_3^\prime|.
\end{align}
Equivalently, one can define the gap-closing condition by the critical momentum $k_*$.
Substituting Eq.~\ref{eq:defalpha} into Eq.~\ref{eq:a2a0ne0} gives
\begin{align}
    \cos k=\frac{J_3J_3'\cos(2\phi)-J_2J_2'}{\alpha_2},
    \label{eq:cosk_phi}
\end{align}
so that for a fixed $\phi$, the gap closes at the critical momenta $k_*= k_*(\phi)$ (mod $2\pi$), where
\begin{align}
    k_*(\phi)=\pm\arccos\!\left[\frac{J_3J_3'\cos(2\phi)-J_2J_2'}{\alpha_2}\right],
    \quad k_*(\phi)\in[0,\pi].
    \label{eq:kstar}
\end{align}
The existence of a real $k_*(\phi)$ requires
\begin{align}
    \left|\frac{J_3J_3'\cos(2\phi)-J_2J_2'}{\alpha_2}\right|\le 1,
    \label{eq:kstar_cond}
\end{align}
which determines the interval of $\phi\in(0,\pi/2)$ for which the spectrum remains gapless.
\par
When $J_{2}^{2}-J_{3}^{2}=J_{2}^{\prime2}-J_{3}^{\prime2}=0$, the gap-closing condition reduces to
\begin{align}
    \cos(2\phi)=\frac{J_2J_2^{\prime}}{J_3J_3^{\prime}}=\pm 1.
\end{align}
leading to $\phi_c=0,~\pi/2~({\mathrm{mod}}~\pi)$. Since we restrict the analysis to $\phi\in(0,\pi/2)$, the system remains gapped in our case, and no topological transition occurs.

\bibliography{biblio}

\end{document}